# Scintillation efficiency of liquid xenon for nuclear recoils with the energy down to 5 keV

V. Chepel[*], V. Solovov, F. Neves, A. Pereira, P.J. Mendes, C.P. Silva, A. Lindote,

J. Pinto da Cunha, M.I. Lopes

*LIP-Coimbra and C.F.R.M. of the Department of Physics, University of Coimbra,*

*3004-516 Coimbra, Portugal*

S. Kossionides

*NCSR DEMOKRITOS, Institute of Nuclear Physics, Gr-15310 Aghia Paraskevi, Greece*

**Abstract.** The scintillation efficiency of liquid xenon for nuclear recoils has been measured to be nearly constant in the recoil energy range from 140 keV down to 5 keV. The average ratio of the efficiency for recoils to that for gamma-rays is found to be 0.19±0.02.

## 1. Introduction

There is good evidence that only a small fraction of the mass of the universe is directly observable. The missing matter is believed to be some form of yet undetected Weakly Interacting Massive Particle (WIMP). Various experiments aim at detecting WIMPs by scattering in matter. In particular, liquid xenon is a suitable detection media [1]. Detecting the low energy recoils resulting from collisions with WIMPs requires knowledge of the scintillation yield due to nuclear recoils of very low energy (~few keV). Here we report the measurement of the scintillation of liquid xenon for nuclear recoils in the range from 5 keV to 140 keV.

---

[*] Corresponding author, e.mail vitaly@lipc.fis.uc.pt

The scintillation efficiency of liquid xenon for nuclear recoils normalized to that for gamma-rays has been measured previously for recoil energies ranging from 40 keV to 150 keV [2-4]. However, the values measured by these experiments are inconsistent; efficiencies about 0.5 [2] and 0.2 [3, 4] were reported. Furthermore, the existing data does not cover the energies below 40 keV, of utmost importance for WIMP search experiments. In the present paper we report on the measurement of the relative scintillation efficiency of liquid xenon for nuclear recoils to 122 keV γ-rays.

## 2. Technique

Xenon recoils were obtained by irradiating a liquid xenon target with monoenergetic fast neutrons (6 MeV and 8 MeV) and selecting events that correspond to elastic scattering at a fixed angle θ. The kinematic equations give the recoil energy as

$$E_R \approx 2E_n \frac{m_n M_{Xe}}{(M_{Xe} + m_n)^2}(1 - \cos\theta) \qquad (1)$$

where $E_n$ is the energy of incident neutrons, and $m_n$ and $M_{Xe}$ are the masses of the neutron and of the target nucleus, respectively, with $M_{Xe} \gg m_n$.

The liquid xenon chamber was designed specifically for this experiment. It comprises a thin wall stainless steel cylinder inserted into an aluminium vessel kept under vacuum for thermal insulation. The active volume of liquid xenon, defined by PTFE reflectors (see Fig.1), has a diameter of 163 mm and 55 mm height (about 1.1 litre). A 2 cm thick dead layer of liquid xenon surrounds the active volume from the lateral side to reduce the low energy gamma-ray background. The active volume is viewed from the top by an array of 7 photomultipier tubes (PMT), with the windows immersed into the liquid, through circular openings made in the PTFE sheet. Each tube has a 2-inch quartz window and bialkali photocathode with a metal pattern underlying the photocathode to reduce its resistivity at low temperature [5]. The PMTs were selected by the manufacturer for a quantum efficiency of >20% for 175 nm at room temperature.

A $^{57}$Co radioactive source, emitting predominantly γ-rays of 122 keV, was placed under the chamber bottom, inside the vacuum insulation, for calibration purposes. The source could be moved from the outside and positioned either in one of 7 fixed positions, each aligned with a photomultiplier, or hidden in a lead castle. Windows of 8 mm diameter 0.5 mm thick were machined in the stainless steel chamber bottom at those positions.

Xenon was purified by passing it through an Oxisorb column. During two weeks of operation at the neutron beam, we noticed no change of the scintillation amplitude. More details on the design of the chamber, its operation and xenon purification are reported in [6].

The chamber has been thoroughly tested with a gamma-ray source as reported in [6]. These bench tests show a high light collection efficiency, yielding in average ≈5.5 photoelectrons per 1 keV of deposited energy (as measured with 122 keV γ-rays). The time resolution was measured with Compton electrons, varying from 1.3 ns to 4.2 ns in the energy range from 120 keV down to 15 keV, respectively. An energy resolution of 18% for 122 keV γ-rays was obtained. All resolutions are given as full width at half maximum (fwhm).

The measurements were carried out with a monochromatic neutron beam at the Demokritos tandem accelerator in Athens. Neutrons were produced by means of (d-d) reactions by irradiation of a deuterium target with accelerated deuterons. The target is a gas cell of 10 mm diameter and 3.7 cm length, with a 5 µm thick molybdenum window. The pressure inside the cell was varied between 1.2 and 2.2 bar, for different runs. This setup provides monoenergetic neutrons with energies of 6 MeV or 8 MeV. The width of the energy distribution is < 1% (standard deviation) in the forward direction.

A simplified layout of the experimental setup is shown in Fig.2. The liquid xenon chamber was placed at a distance of 3 m from the deuterium target. The neutron beam was collimated with a paraffin cylindrical collimator placed right in front of the deuterium cell. Paraffin blocks, some of them borated, around the gas target provided additional shielding. Further blocks were placed in the floor between the target and the chamber, in order to reduce the background due to the neutrons scattered or captured in the surrounding materials.

To detect neutrons that scatter in the liquid xenon chamber we used a NE213 neutron detector followed by a pulse shape n/γ discriminator set to γ-ray pulse suppression mode [7]). The detector was placed at about 1 m from the chamber and at various angles, θ, relative to the incident beam. The NE213 detector was placed inside a paraffin cylinder (see Fig.2). The measurements refer to incident neutron energies of 8 MeV and 6 MeV and an angle θ that was varied between 20º and 100º. Neutrons scattered at these angles produce xenon recoils with the energies in the range from 140 keV down to 5 keV. The incident neutron flux density was estimated to be ~$10^5$ neutrons/(s·cm$^2$).

The front-end electronics and the data acquisition system are described elsewhere [6]. The data acquisition was triggered by the PMT signals from the liquid xenon chamber, if two or more PMTs showed a signal above the threshold set at ≈ 1 photoelectron. Additionally, a signal from the neutron detector was required within 300 ns. The signal amplitude and the arrival time of each of 7 photomultipliers, as well as the arrival time for the signal from the neutron detector, were recorded for each event. The delay between the trigger generated by the liquid scintillation and the signal from the neutron detector (herein called time-of-flight, TOF) provided the time of propagation of a scattered particle between the chamber and the neutron detector.

**3. Results**

An example of the time-of-flight distribution is shown in Fig.3a. The left most peak is due to γ-rays that undergo Compton scattering in the chamber and hit the neutron detector. These correspond to NE213 signals that are not rejected by the pulse shape discriminator. The second peak, at about 21 ns after that due to γ-rays, corresponds to neutron elastic scattering events within the active volume. The broad distribution at larger times corresponds to the events that involve at least one inelastic interaction in the chamber. By setting a proper time of flight window, we can select events that correspond to elastic scatterings in the liquid xenon.

The photomultipliers were periodically calibrated between runs, *in situ*, at the operating conditions with a pulsed light emission diode [6]. The number of photoelectrons emitted by the photocathode is inferred from the statistical analysis of the amplitude distribution (method 1 in [8]). The total amplitude of the scintillation signal in liquid xenon is obtained offline as the sum of the amplitudes of all 7 photomultiplier tubes expressed in number of photoelectrons. Fig.3b shows an example of the pulse height distribution of the scintillation signals due to 33 keV xenon recoils, measured with neutrons of 6 MeV at the scattering angle of 50º.

To obtain the scintillation efficiency for nuclear recoils we first determined the most probable value of the amplitude distribution of the signals from the elastic scattering. This is then divided by the recoil energy, as given by Eq.1, assuming that the collisions take place at the centre of the liquid volume. The ratio was then normalized by the corresponding value obtained with 122 keV gamma-rays from the $^{57}$Co source positioned under the chamber bottom, at the axis. The relative scintillation efficiency obtained in this way is shown in Fig.4 as a function of the recoil energy.

The relative scintillation yield appears to be practically constant for recoil energies above 8 keV, the average being 0.19±0.02, probably with some increase at lower energy. In the range from 40 keV to 140 keV, it is in a good agreement with previously published results [3, 4] but far from the value referred in [2]. The general trend is similar to that observed for iodine recoils in NaI(Tl), where an increase of the efficiency is also seen at recoil energies below ≈20 keV [9, 10], and seems to disagree with theoretical estimates [11, 12]. An increase of efficiency at low energy was also observed for Na recoils in NaI(Tl) and Ca and F recoils in CaF$_2$ scintillation crystals [10].

The uncertainties shown in Fig.4 are estimated from the geometrical acceptance and are determined by the minimum and the maximum scattering angles measured in the horizontal plane that are possible for each angular position of the neutron detector, hence, these error bars are somehow pessimistic. The number of events is typically ≈800 per point in Fig.4. Thus, the statistical error in determining the peak position in the pulse height spectrum is small and does not exceed 5%.

**3. Discussion**

Among the factors, which may contribute to the systematics of our measurements, we identify: i) the neutron induced gamma background, ii) the contribution from the events involving inelastic scattering of neutrons both within the active xenon and/or in the surrounding materials, and iii) multiple elastic scattering events in the chamber from which at least one interaction occurs in the active volume.

The neutron induced gamma background has been substantially suppressed by using thick paraffin shielding surrounding the collimator and the NE213 neutron detector with a pulse shape n/γ discriminator. In addition, a 2 cm passive liquid xenon layer surrounding the active volume also works as an absorber for low energy γ-rays originating outside the chamber. The residual gamma-background was efficiently cut by selecting a proper TOF window (see Fig.3a). Moreover, the γ-rays that interact in the chamber deposit normally a large amount of energy as compared to the energy expected from nuclear recoils, so that most of the signals due to γ-rays fall out of the energy scale set in the acquisition system and are rejected.

As for the inelastic scattering, the neutron usually transfers a significant energy to the nucleus both to its excitation and recoil. The energy of the outgoing neutron is therefore smaller than that of the elastically scattered neutrons and their time-of-flight is larger by about 20 ns (see Fig.3a). Moreover, the de-excitation gamma-rays scatter in the liquid xenon with a probability of about 0.96 (obtained by Monte Carlo simulation). Hence, the energy deposition is expected to be much higher than for elastic recoils. After the cuts on TOF and on the signal amplitude the contribution from inelastic events is estimated to be less than 0.5%. These events are due to nuclear recoils resulting from neutron inelastic scattering followed by the escape of γ-rays.

A Monte Carlo simulation has also been used to assess the contribution from multiple elastic scattering. Fig.5 shows examples of the distribution of the energy deposited in the active volume of the chamber for events that involve at least one interaction in the active region (the other(s) occurring either in the active volume, in the dead layer of xenon or in the chamber walls). Both the relative contribution and the energy spectrum of the multiple events vary with the angle, θ, at which the neutron detector is set (see Fig.2). This reflects the angular dependence of the elastic cross section. In fact, the elastic cross section has maxima at 0º, 57º and 101º. Thus, for angles θ close to those values, the contribution from multiple scattering events is smaller. Fig.5 shows the simulated spectra of the energy deposited in the active volume of the chamber for different types of events for a) θ=20º and b) θ=57º. For 20º, the fraction of single scattering events is about 89% in the energy range 0 keV to 10 keV, whereas for θ=57º it is 83% in the energy range 22 keV to 60 keV. The remaining events are the multiple scattering events with at least one interaction in the sensitive volume. As seen from Fig.5, these events are broadly distributed. Moreover, the mean value of the deposited energy for these events practically coincides with that due to single scattering events, for these angles. The difference between the average energy deposited in the single elastic scattering events and that for the single plus multiple scattering events is estimated to be less than 0.5%.

Fig.5c displays the simulated pulse height spectrum for θ=45º. The elastic cross section has a minimum near this angle and, therefore, a larger contribution from multiple scattering is seen. In this case, the single scattering events are about 68% of the total in the energy range 12 keV to 50 keV. The energy distribution of multiple scattering events is shifted to lower values resulting in a slowly increasing background at low energies. However, they shift the mean deposited energy by about 0.3 keV, only. This example corresponds to the largest contribution from multiple events amongst all our measurements.

Thus, and taking into account that the relative scintillation efficiency practically does not depend on the recoil energy (Fig.4), we conclude that multiple scattering can be neglected in our results.

The comprehensive analysis of the uncertainties and systematic errors is only possible on the basis of a detailed Monte Carlo simulation, which includes hadronic and electromagnetic processes as well as the propagation of scintillation light in the liquid xenon chamber. Such simulation is being developed using GEANT4 [13]. Some preliminary results were already referred in the above discussion. Furthermore, we expect to obtain a detailed Monte Carlo description of the observed pulse height spectra at our experimental conditions so that the contributions from different processes could be identified and subtracted (although it is already clear that these are small corrections). In particular, we believe that the uncertainties associated both to the recoil energy and to the scintillation efficiency can be substantially reduced. Additionally, a detailed mapping of the light collection within the chamber should allow reconstruction of the interaction position. Preliminary results obtained with γ-rays are reported in [6] and [14]. By reconstructing the interaction position within the liquid xenon chamber, the scattering angle definition should be significantly improved. Moreover, some of the remaining multiple scattering events might be identified and rejected.

**Acknowledgements**

The authors wish to thank the personnel of the Demokritos tandem accelerator for their help in running the experiment in the neutron beam. This work was done under the project NoPOCTI/FP/FNU/50208/2003 from Fundação para a Ciência e Tecnologia, Portugal. V.Solovov, F.Neves, A.Lindote and C.P.Silva acknowledge the support from SFRH/BPD/14517/2003, SFRH/BD/3066/2000, SFRH/BD/12843/2003 and SFRH/BD/19036/2004 fellowships from the same institution, respectively.

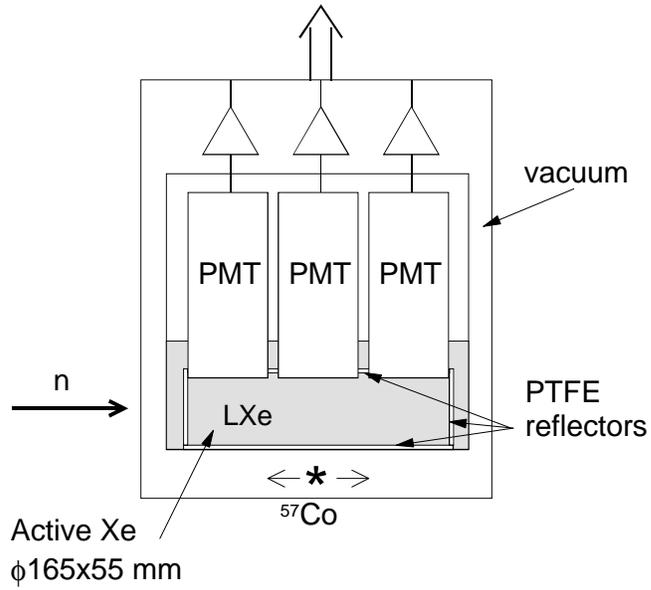

Figure 1 Schematic drawing of the liquid xenon chamber.

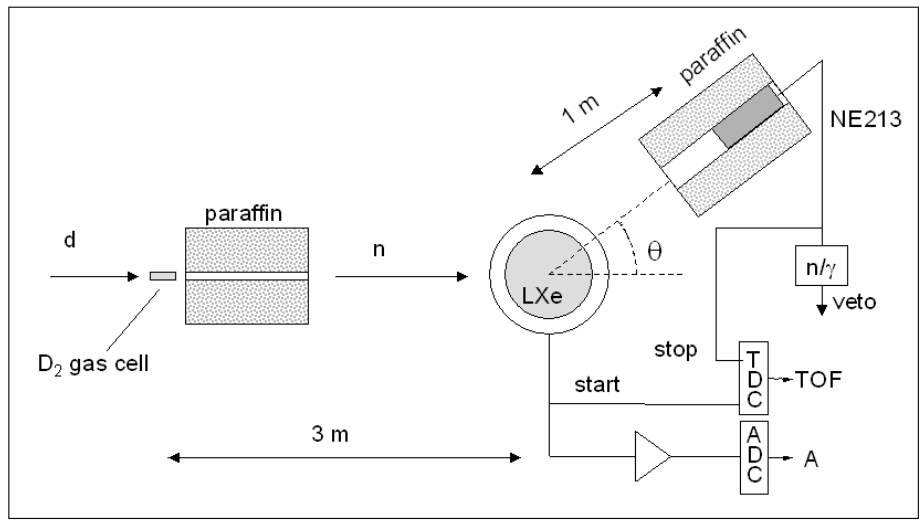

Figure 2 Layout of the experimental setup. Two neutron energies were used, 6 MeV and 8 MeV; the angle θ was varied between 20º and 110º.

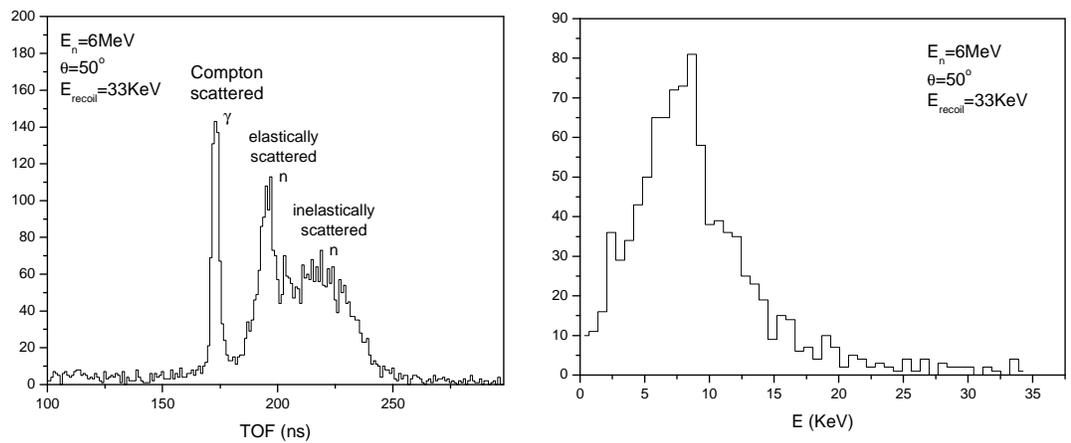

Figure 3 Examples of time-of-flight spectrum (a) and pulse height distribution of the signals corresponding to xenon recoils of 33 keV energy (b) The energy scale is calibrated with 122 keV gamma-rays. The recoil energy is calculated assuming that the interactions occur at the centre of the chamber.

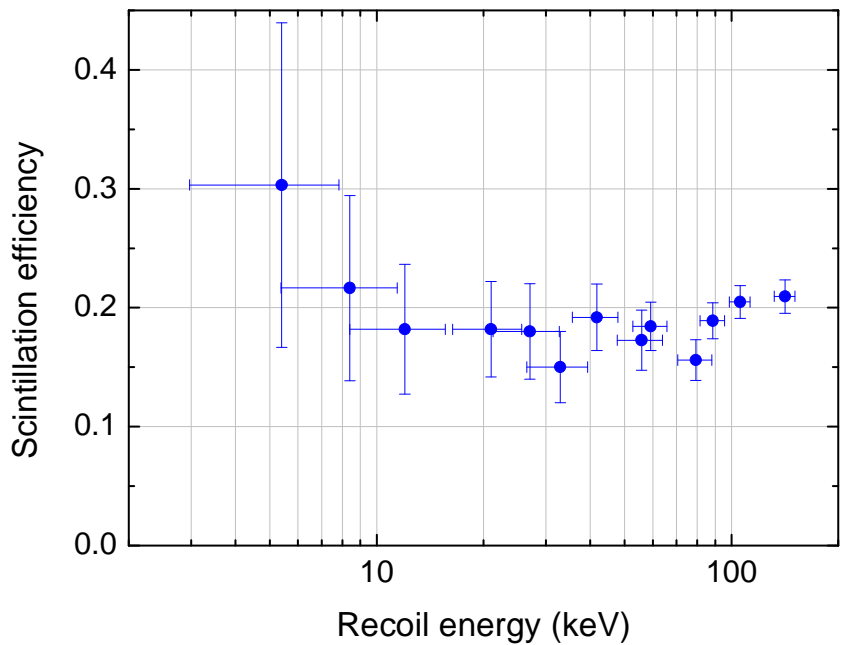

Figure 4 Scintillation efficiency of liquid xenon for nuclear recoils normalized to that for 122 keV γ-rays.

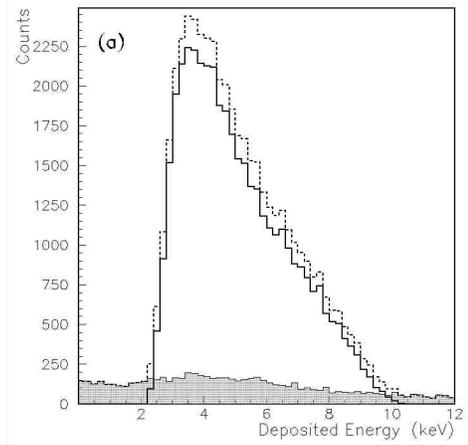
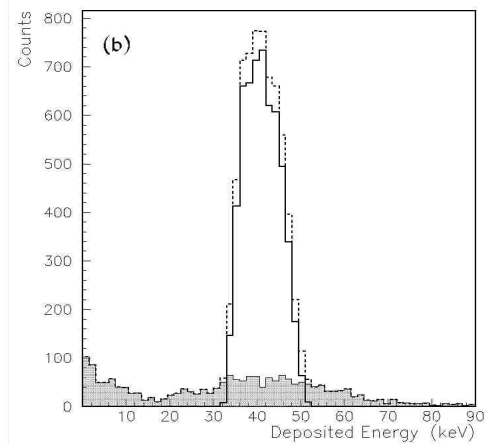
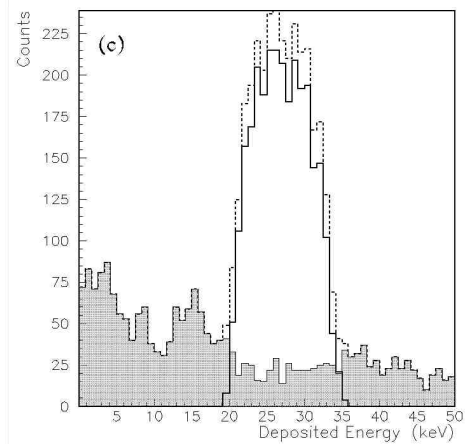

Figure 5 Simulated distributions of the deposited energy in the sensitive volume due to single and multiple elastic scattering events for 6 MeV incident neutrons at (a) θ=20º, (b) θ=57º and (c) θ=45º. The solid line corresponds to single scattering events; shadowed area shows the multiple scattering events with at least one interaction in the sensitive volume; the dashed line is the sum of the two.